\documentclass[preprint,prb,showpacs,preprintnumbers,amsmath,amssymb]{revtex4}

\usepackage{graphicx}
\usepackage{dcolumn}
\usepackage{bm}

\usepackage{colordvi}

\begin{document}

\title{Polyelectrolyte stars in planar confinement}

\author{Martin Konieczny}
  \email{kon@thphy.uni-duesseldorf.de}
\author{Christos N.\ Likos}
\affiliation{
Institut f\"ur Theoretische Physik II, 
Heinrich-Heine-Universit\"at D\"usseldorf, 
Universit\"atsstra{\ss}e 1, 
D-40225 D\"usseldorf, Germany}

\date{{\bf April 11, 2006}, submitted to 
{\sl Journal of Chemical Physics}}

\begin{abstract}
We employ monomer-resolved Molecular Dynamics simulations and theoretical
considerations to analyze the conformations of multiarm polyelectrolyte
stars close to planar, uncharged walls. We identify three mechanisms that
contribute to the emergence of a repulsive star-wall force, namely:
the confinement of the counterions that are trapped in the star interior,
the increase in electrostatic energy due to confinement as well as 
a novel mechanism arising from the compression of the stiff polyelectrolyte
rods approaching the wall. The latter is not present in the case of
interaction between two polyelectrolyte stars and is a direct consequence
of the impenetrable character of the planar wall. 
\end{abstract}

\pacs{82.70.-y, 82.35.Rs, 61.20.-p}

\maketitle


\section{Introduction}

Polyelectrolytes (PE's) are polymer chains that carry ionizable groups
along their backbones. Upon solution in aqueous solvents, these 
groups dissociate, leaving behind a charged chain and the corresponding
counterions in the solution. Chemically anchoring $f$ such PE-chains
on a colloidal particle of radius $R_{\rm d}$ gives rise to a spherical
polyelectrolyte brush (SPB). In the limit in which the height $R$ of
the brush greatly exceeds $R_{\rm d}$, one talks instead about polyelectrolyte 
stars. A great deal of theoretical\cite{pincus:91,klein:macromolecules:02,klein:macromolecules:99,borisov:jp:97,borisov:epjb:98,linse:macrom:00,arben:prl:02,arben:jcp:02,denton:pre:03,wang:denton:pre:04,norm:jcp:04,arben:cps:04,roger:epje:02,borisov:macrom:02,zhulina:macrom:02,borisov:macrom:03,arben:jcp:06} and 
experimental\cite{guenoun:prl:98,maarel:macrom1:00,maarel:macrom2:00,maarel:langmuir:00,abraham:langmuir:00,ballauff:macrom:00,guo:pre:01,nico:02,heinrich:epje:01,maarel:prl:05}
effort has been
devoted in the recent past with the goal of understanding the 
conformations and the interactions of SPB's and PE-stars. The reasons
are many-fold. The grafted PE-chains can provide an electrosteric
barrier against flocculation of the colloidal particles on which
the chains are grafted, rendering the systems very interesting from
the point of view of colloidal stabilization. 
Moreover, they can act as control agents for gelation,
lubrication and flow behavior. From the point of view of fundamental
research, PE-stars are a novel system of colloidal particles that combine
aspects from two different parts of colloid science: polymer theory and
the theory of charged suspensions. Accordingly, the derivation of
the effective interaction between PE-stars has led to their
description as ultrasoft colloids and to theoretical predictions
on their structural and phase behavior\cite{arben:jcp:02,norm:jcp:04} 
that have received already
partial experimental confirmation.\cite{ishizu:mm:05}
Moreover, certain similarities in
the structural and phase behavior of PE-stars and ionic microgels
have been established, demonstrating the close relationship between
the two systems.\cite{dieter:prl:04,dieter:jcp:05}

By and large, the theoretical investigations involving PE-stars have been
limited to the study of either single PE-stars or bulk solutions of the
same. However, a new field of promising applications and
intriguing physics is arising when PE-stars or SPB's are mixed with
hard colloids or brought in contact with planar walls. Indeed,  
PE-stars can be used to model cell adhesion and are also efficient
drug-delivery and protein-encapsulation and 
immobilization 
agents.\cite{riess:pps:03,bhatia:cocis:01,kakizawa:addr:02,biesh:jpcb:05} 
On the other hand, hydrogels,
which are physically similar to PE-stars, adsorbed on planar
walls, form arrays of dynamically tunable, photoswitchable or
bioresponsive 
microlenses.\cite{kim:jacs:04,serpe:am:04,kim:ac:05,kim:jacs:05}
The encapsulation properties 
as well as the characteristics of the microlenses depend sensitively
on the interactions between the PE-stars and the colloidal particles
or the wall, respectively. Therefore, there exists a need to undertake
a systematic effort in trying to understand these interactions 
physically and make quantitative predictions about the ways to
influence them externally. In this paper, we take a first step
in this direction by considering a PE-star in the neighborhood of 
a planar, uncharged and purely repulsive wall. We analyze the
mechanisms that give rise to an effective star-wall repulsion and
identify the counterion entropy and the chain compression against the
planar wall as the major factors contributing to this force. Our
findings are corroborated by comparisons with Molecular Dynamics
simulation results for a wide range of parameters 
that structurally characterize the stars. This work provides
the foundation for examining effects of wall curvature and charge. 

Our paper is organized as follows: Sec.\ \ref{sec:md} is devoted to the
description of the system we are investigating, the simulation
model, and the simulation techniques used. Moreover, we specify the 
physical quantities of interest. Our theoretical approach is addressed 
in detail in Sec.\ \ref{sec:theory}. In Sec.\ \ref{sec:comparison}, we
quantitatively compare and discuss the respective results. Finally, we
conclude and present an outlook to possible future work in Sec.\
\ref{sec:conclusion}.

\section{Simulation model}

\label{sec:md}

We start with a definition of the system under consideration including its 
relevant parameters and a description of the simulation model used. We study 
a dilute, salt-free solution of PE-stars confined between 
two hard walls parallel to the $x$-$y$-plane at positions $z=\pm \tau/2$, 
resulting in an overall wall-to-wall separation $\tau$. We assume a good 
solvent that is only implicitly taken into account via its relative dielectric 
permittivity $\epsilon\approx 80$, i.e., we 
are dealing with an aqueous solution. 
To avoid the appearance of image charges,\cite{torrie:jcp:76,messina:jcp:117}
we assume the dielectric constants to be the same on both sides of the
respective confining walls. It turns out, however, that this is not
a severe assumption, because the 
effective charge of a PE-star is drastically 
reduced compared to its bare value, due to the strong absorption of  
neutralizing
counterions.\cite{klein:macromolecules:99,arben:prl:02,arben:jcp:02,
konieczny:jcp:121} Hence, 
the influence of image charges can be expected to be of minor importance.

The PE-stars themselves consist of $f$ PE-chains, all attached to a common 
colloidal core of radius $R_{\rm d}$, whose size equals the monomer size and is
therefore much smaller than the typical center-to-end length of the chains for
all parameter combinations. The introduction of such a core particle is necessary 
to place the arms in the vicinity of the center, where the monomer density can 
take very high values. The theoretical approach pertains to the limit
of vanishingly small core size. In order to remove effects arising from
the small (but finite) value of the core in the simulation model and
to provide a comparison with theory, we will henceforth employ a consistent
small shift of the simulation data by $R_{\rm d}$. This standard approach
has been applied already, e.g., to the measurement of the interactions
between neutral \cite{jusufi:macromolecules:32} and charged stars \cite{arben:prl:02,arben:jcp:02}, 
in order to isolate the direct core-core interaction effects from the scaling laws valid
at small interstar separations.

The PE-chains are modeled as bead-spring chains of $N$ Lennard-Jones (LJ) 
particles. This approach was first used in investigations of neutral polymer
chains and stars
\cite{stevens:prl:71,grest:macromolecules:20,grest:macromolecules:27}
and turned out to be reasonable. The method was also already successfully
applied in the case of polymer-colloid mixtures\cite{jusufi:jpcm:13}
or polyelectrolyte 
systems.\cite{arben:prl:02,arben:jcp:02,konieczny:jcp:121}
To mimic the above mentioned good solvent conditions, a shifted and truncated
LJ potential is introduced to depict the purely repulsive excluded volume
interaction between the monomers:
\begin{equation}
\label{eq:pot:lj}
V_{\rm LJ}(r)=
\begin{cases}
4\varepsilon_{\rm LJ}\left[\left(\frac{\sigma_{\rm LJ}}{r}\right)^{12}
-\left(\frac{\sigma_{\rm LJ}}{r}\right)^{6}
+\frac{1}{4}\right]&r\leq 2^{1/6}\sigma_{\rm LJ}\\
0&r>2^{1/6}\sigma_{\rm LJ}.
\end{cases}
\end{equation}
Here, $r$ is the spatial distance of two interacting particles. The quantities
$\sigma_{\rm LJ}$ and $\varepsilon_{\rm LJ}$ set the basic length and energy
scales for the system, respectively. In what follows, we fix the system's
temperature to the value $T=1.2\varepsilon_{\rm LJ}/k_{\rm B}$, where 
$k_{\rm B}$ denotes Boltzmann's constant. The PE-chain connectivity is 
modeled by employing a standard finite extension nonlinear elastic (FENE) 
potential:\cite{grest:macromolecules:20,grest:macromolecules:27}
\begin{equation}
\label{eq:pot:fene}
V_{\rm FENE}(r)=
\begin{cases}
-\frac{k_{\rm FENE}}{2}\left(\frac{R_0}{\sigma_{\rm LJ}}\right)^2\ln
\left[1-\left(\frac{r}{R_0}\right)^2\right]&r\leq R_0\\
\infty&r>R_0,
\end{cases}
\end{equation}
with a spring constant $k_{\rm FENE}=7.0\varepsilon_{\rm LJ}$. The divergence
length $R_0$ limits the maximum relative displacement of two neighboring
monomers and is set to $R_0=2.0\sigma_{\rm LJ}$ in the scope of this work.
The described set of parameters determines the equilibrium bond length, 
in our case resulting in a value $l_0=0.97\sigma_{\rm LJ}$.

When modeling the interactions between the monomers and a star's colloidal
core, the finite radius $R_{\rm d}$ of the latter has to be taken into account.
All monomers experience a repulsive interaction with the central particle, in 
analogy to Eq.\ (\ref{eq:pot:lj}) reading as
\begin{equation}
\label{eq:pot:c}
V^{\rm c}_{\rm LJ}=
\begin{cases}
\infty&r\leq R_{\rm d}\\
V_{\rm LJ}(r-R_{\rm d})&r>R_{\rm d}.
\end{cases}
\end{equation}
In addition, there is an attraction between the innermost monomers in the 
arms' chain sequence and the core which is of FENE type and can be written
as (cf.\ Eq.\ (\ref{eq:pot:fene})):
\begin{equation}
V^{\rm c}_{\rm FENE}=
\begin{cases}
\infty&r\leq R_{\rm d}\\
V_{\rm FENE}(r-R_{\rm d})&r>R_{\rm d}.
\end{cases}
\end{equation}

The chains are charged in a periodic fashion by a fraction $\alpha$ in such 
a way that every $(1/\alpha)$-th monomer carries a monovalent charge.
Consequently, there is a total number of $N_c=\alpha fN$ monomer ions per 
PE-star. To ensure electroneutrality of the system as a whole, we include the
same amount of oppositely charged, monovalent counterions in our 
considerations. Since the latter are able to freely move, they have to be 
simulated explicitly. Furthermore, they are of particular importance because
they are expected to crucially affect the physics of the system. One example
is the aforementioned fact that they induce a reduction of the stars' bare
total charges.

Two charged beads with spatial distance $r_{ij}$ interact by a full 
Coulomb potential, i.e., the electrostatic interaction energy is
\begin{equation}
\label{eq:pot:coul}
\beta V_{\rm Coul}(r_{ij})=
\beta \frac{q_i q_j e^2}{\epsilon r_{ij}}
\equiv\lambda_{\rm B}\frac{q_i q_j}{r_{ij}}.
\end{equation}
Thereby, $q_i,q_j=\pm 1$ are the valencies of monomer ions and counterions,
respectively, $e$ denotes the absolute value of the unit charge, and 
$\beta=1/k_{\rm B}T$ is the inverse temperature. In the above equation, 
the so-called Bjerrum length
\begin{equation}
\lambda_{\rm B}=\frac{\beta e^2}{\epsilon}
\end{equation}
was introduced. It is defined as the distance at which the electrostatic
energy equals the thermal energy. Thus, it characterizes the interaction
strength of the Coulomb coupling. In the case of water at room temperature,
one obtains $\lambda_{\rm B}=7.1\,{\rm {\AA}}$. 
In our simulations, we fix the Bjerrum
length to $\lambda=3.0\sigma_{\rm LJ}$, thus corresponding to an experimental
particle diameter $\sigma=2.4\,{\rm {\AA}}$. 
This is a realistic value for typical
polyelectrolytes.\cite{essafi:jphys2f:5}

For the purpose of completing the set of interaction potentials needed
to describe the system at hand, we have to define particle-wall interactions.
For technical reasons (see below), we do not regard the walls as true hard 
walls, as common in Monte-Carlo (MC) simulation 
studies.\cite{messina:langmuir:19,messina:macromolecules:37}
Following the course of our above modeling and based on 
Eq.\ (\ref{eq:pot:lj}), we assume them to be of 
truncated-and-shifted-LJ type instead, leading to 
the following monomer-wall interaction:
\begin{equation}
\label{eq:pot:w}
V^{\rm w}_{\rm LJ}(z)=V_{\rm LJ}\left(\frac{\tau}{2}-z\right)
+V_{\rm LJ}\left(\frac{\tau}{2}+z\right),
\end{equation}
whereas $z$ refers to the $z$-component of the position vector of the 
particular bead. Likewise, the potential function for a star's core 
interacting with the confining walls is yielded by combining Eqs.\ 
(\ref{eq:pot:c}) and (\ref{eq:pot:w}):
\begin{equation}
V^{\rm wc}_{\rm LJ}(z)=V_{\rm LJ}\left(\frac{\tau}{2}-R_{\rm d}-z\right)
+V_{\rm LJ}\left(\frac{\tau}{2}-R_{\rm d}+z\right).
\end{equation}

To finalize this section, we present a short summary of the 
simulation techniques used. We perform monomer-resolved Molecular 
Dynamics (MD) simulations in the canonical ensemble, employing a 
rectangular simulation box of total volume $V=M^2\tau$ that contains a 
single PE-star. We apply periodic boundary conditions in the $x$- and 
$y$-directions, while the box is confined with respect to the $z$-direction. 
Here, we always fix $M=\tau=120\sigma_{\rm LJ}$. 
This provides a sufficiently large simulation box to suppress 
any undesirable side-effects {\it a priori} and emulates a dilute solution
of PE-stars (cf.\ above description of the physical problem under 
investigation). For the numerical integration of the equations of motion,
we adopt a so-called Verlet algorithm in its velocity 
form.\cite{allen:book,frenkel:book,rapaport:book} 
In order to stabilize the system's temperature, we make use of a
Langevin thermostat\cite{allen:book,frenkel:book,rapaport:book} that 
introduces additional friction and random forces with appropriately
balanced, temperature-dependent amplitudes.
Due to the periodic boundary conditions and the long-range character of 
the Coulombic forces, a straight-forward calculation of the latter 
pursuant to Eq.\ (\ref{eq:pot:coul}) is not feasible. Therefore, we 
have to evaluate the forces using Lekner's well-established summation 
method\cite{lekner:physicaa:176} in its version for quasi two-dimensional 
geometries. Thereby, the convergence properties of the sums occurring 
during the computation are enhanced by a mathematically accurate rewriting, 
allowing a proper cut-off. For performance reasons, the forces
have to be tabulated.

The typical time step is $\Delta t=0.002t_0$, with 
$t_0=(m\sigma_{\rm LJ}^2/\varepsilon_{\rm LJ})^{1/2}$ being the
associated time unit and $m$ the monomer mass. In our case, the counterions
are taken to have the same mass and size as the (charged) monomers.

In our simulations, we measure the effective core-wall forces as a function 
of the center-to-wall distance $D$. Furthermore, we examine various
other static quantities and their $D$-dependence, namely the 
center-to-end distance $R$, the radius of gyration $R_{\rm g}$,
the number of condensed counterions and the density profiles of 
all particle species involved. For this purpose, for every particular
value of $D$ the system is equilibrated for about
$5\times 10^5$ time steps. After this equilibration period, we perform
production runs lasting between $1\times 10^6-2\times 10^6$ time steps.
We carry out the above described measurements for a variety of arm 
numbers $f$ ($f=$10, 18, 30) and charging fractions $\alpha$ 
($\alpha=$1/5, 1/4, 1/3) in order to make systematic predictions for the
$f$- and $\alpha$-dependencies of all theoretical parameters. In all cases,
we fix the degree of polymerization, i.e., the
number $N$ of monomers per arm, to a value
$N=50$.

\section{Theory of the effective PE-star--wall interaction}
\label{sec:theory}

A PE-star in the neighborhood of a planar, impenetrable wall, undergoes
conformational changes that modify its (free) energy in comparison to
the value it has when the wall is absent or far away from the star.
This separation-dependent difference between the free energies is
precisely the effective interaction between the PE-star and the wall.
There are three distinct mechanisms that give rise to an effective
interaction in our case: the change in the electrostatic energy of the
star, the change in the counterion entropy arising from the presence
of the geometric confinement as well as contributions from compression
of the stiff PE-chains against the flat wall. The former two are
intricately related to each other, as the counterion distribution is
dictated by the strong Coulomb interactions, hence they will be
examined jointly, the latter is a distinct phenomenon arising
in the presence of impenetrable walls. 

\subsection{The electrostatic and entropic contributions}

\label{subsec:ee}

\begin{figure}[h]
\begin{center}
\includegraphics[width=8cm,draft=false,clip]{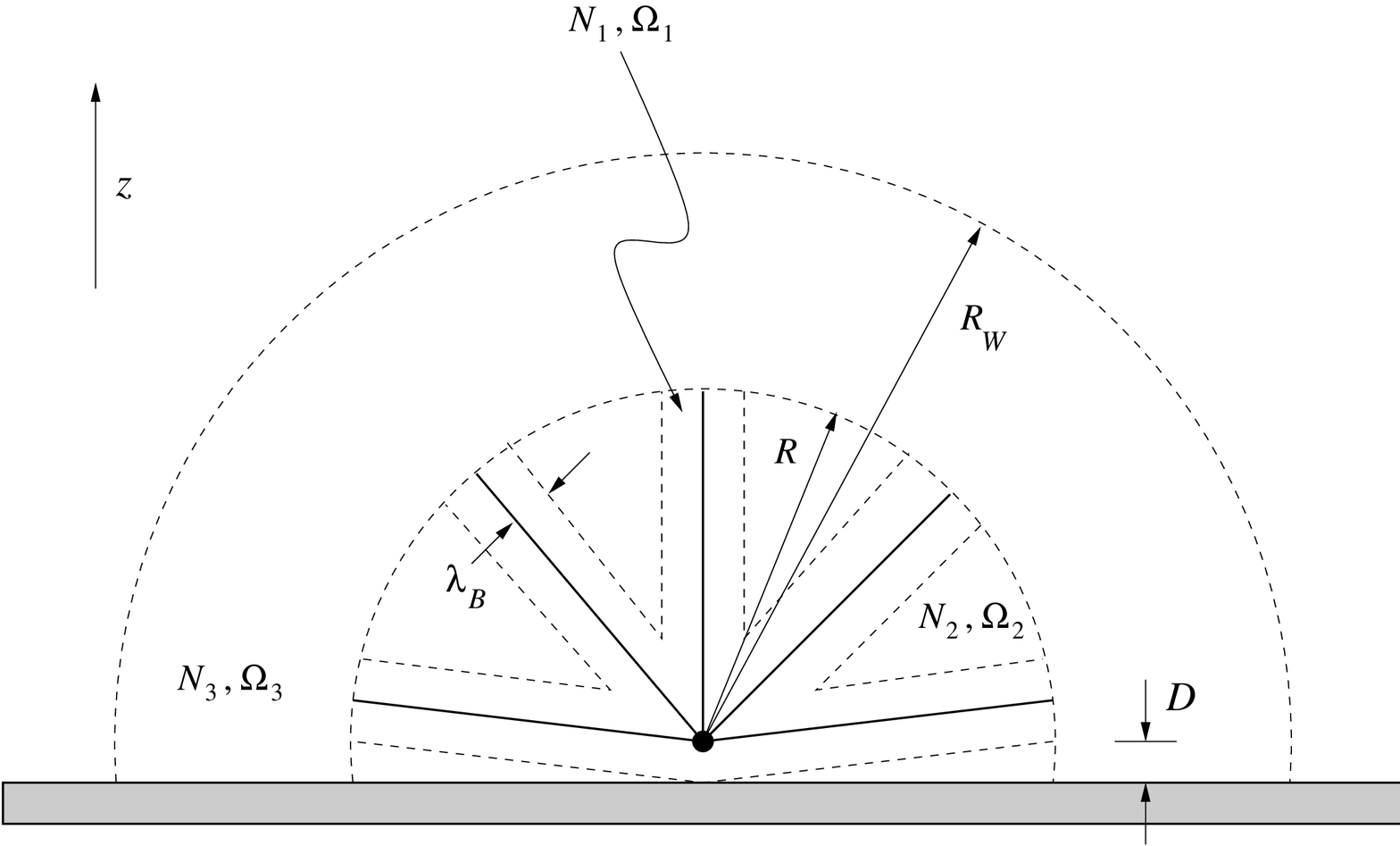}
\end{center}
\vspace{-0.5cm}
\caption{
\label{fig:sketch}
Konieczny, Likos
}
\end{figure}

To obtain theoretical predictions for the electrostatic-entropic 
contribution to the free energy for a
PE-star close to a flat, hard wall, 
we employ a mean-field approach, inspired by 
and akin to that developed
in Refs.\ \onlinecite{klein:macromolecules:99,arben:prl:02,arben:jcp:02},
and \onlinecite{konieczny:jcp:121}.
Let $\rho_{\rm s}=N_{\rm s}/\Omega$ be the density of
the solution of PE-stars, where $N_{\rm s}$ denotes the 
total number of PE-stars in the macroscopic volume $\Omega$. In general, a 
single PE-star of total charge $Q$ is envisioned as a spherical object of 
radius $R$, embedded in a spherical Wigner-Seitz cell 
of radius $R_{\rm W}$. Additionally, the latter contains $N_{\rm c}$ 
counterions, restricted to move in the Wigner-Seitz cell only and forming an 
oppositely charged background of total charge $-Q$. The cell's radius is 
connected to the star density via  $R_{\rm W}=(4\pi\rho_{\rm s}/3)^{1/3}$. 
In the case of center-to-wall separations smaller than $R_{\rm W}$ and/or $R$, 
we simply cut the counterion cloud and/or the star itself at the confining
wall and treat them as chopped spheres instead of full spheres. 
Obviously, the compound system composed of 
a PE-star and its counterions is electroneutral as a whole. 
Since we are interested in the case of dilute PE-solutions only, we 
can limit ourselves to the consideration of a single PE-star.
Clearly, for center-to-wall 
distances $D>R_{\rm W}$ there is no interaction between a PE-star 
and the wall within the framework of our theory. Accordingly, we will deal
with the case $D\leq R_{\rm W}$ only. Fig.~\ref{fig:sketch} 
sketches the physical situation and visualizes the decisive length scales
of the problem.

A mechanism crucially influencing the physics of PE-systems is the so-called
Manning condensation of counterions.\cite{manning:jcp:51-3a,manning:jcp:51-3b,
manning:jcp:51-8,odijk:jpolymsci:16,winkler:prl:81,nyquist:macromolecules:32,
jiang:jcp:110,deserno:macromolecules:33} A parameter indicating whether or
not such a condensation effect will occur is the dimensionless
fraction $\xi=\lambda_{\rm B}N\alpha/R$. If it exceeds unity, counterions
will condense on the arms of the 
stars.\cite{manning:jcp:51-3a,manning:jcp:51-3b,manning:jcp:51-8} In our 
case, this condition is true for all parameter combinations examined (see
below). Hence, the condensation effect has to be taken into account in our 
theoretical modeling. For that reason, we partition the counterions in 
three different states, an ansatz already put forward in 
Refs.\ \onlinecite{arben:prl:02,arben:jcp:02}, and
\onlinecite{kramarenko:mts:9}: $N_1$ of the $N_{\rm c}$ counterion 
are in the condensed state, i.e., they are confined in imaginary hollow 
tubes of outer radius $\lambda_{\rm B}$ and inner radius $\sigma_{\rm LJ}$ 
around the arms of the star. The total volume\footnote{
Here, the tube overlap in the vicinity of the colloidal core is not 
taken into account. Accordingly, the resulting reduction of the volume 
$\Omega_1$ with respect to the value given in the main text is neglected. 
Anyway, this simplification does not considerably affect the theoretical 
results.} accessible for the counterions in this state
is $\Omega_1=\pi(\lambda_{\rm B}^2-\sigma_{\rm LJ}^2)Rf$. A number of $N_2$ 
counterions is considered to be trapped in the interior of the star. 
These ions are allowed to explore the overall volume 
$\Omega_{\rm in}=\Omega_{\rm cs}(D,R)$ of the (possibly chopped) star except 
the tubes introduced above, i.e., $\Omega_2=\Omega_{\rm cs}(D,R)-\Omega_1$. 
Thereby, the volume $\Omega_{\rm cs}$ of a chopped sphere can be derived 
by straightforward calculations. With $\Omega(R)=4\pi R^3/3$, one obtains:
\begin{equation}
\frac{\Omega_{\rm cs}(D,R)}{\Omega(R)}=
\begin{cases}
\frac{1}{2}+\frac{3}{4}
\left(\frac{D}{R}\right)
-\frac{1}{4}\left(\frac{D}{R}\right)^3&D\leq R\\
1&D > R.
\end{cases}
\end{equation}
Note that
the total volume of the star is big compared to the volume of the tubes
surrounding the arms, $\Omega_{\rm cs}(D,R)\gg \Omega_1$. Thus, we will make
use of the approximation $\Omega_2\approx \Omega_{\rm cs}(D,R)$ in all
steps to come. The remaining $N_3=N_{\rm c}-N_1-N_2$ counterions are able 
to freely move within the outer shell of volume 
$\Omega_{\rm out}=\Omega_3=\Omega_{\rm cs}(D,R_{\rm W})-\Omega_{\rm cs}(D,R)$ 
surrounding the star, where $\Omega_{\rm cs}(D,R_{\rm W})$ is 
the total volume of the 
Wigner-Seitz cell. The subdivision of counterions in
the three different states is also depicted in Fig.\ \ref{fig:sketch},
showing a star with 5 arms that are assumed to be fully stretched for 
demonstration.

The electrostatic and entropic part $V_{\rm ee}(D)$ of the
effective interaction $V_{\rm eff}(D)$ between the PE-star
and the wall, kept at center-to-wall distance $D$, results after
taking a canonical trace over all but the star center degree of freedom.
With ${\cal F}_{\rm ee}(D;R,\{N_i\})$ being the variational Helmholtz free 
energy for a system where a single star faces the confining wall, 
and which contains just the electrostatic and entropic parts,
it is 
defined as:\cite{likos:physrep:348}
\begin{equation}
\label{eq:potential}
V_{\rm ee}(D)=\min_{R,\{N_i\}}{\cal F}_{\rm ee}(D;R,\{N_i\}).
\end{equation}
In principle, the equilibrium values of $R$ and $\{N_i\}$ are 
determined by the above minimization. We will discuss this point in
more detail shortly. Note that we neglected a second, 
$D$-independent  term on the right-hand side of 
Eq.\ (\ref{eq:potential}) representing the contribution to the free 
energy for an infinitely large center-to-wall separation. Since it makes 
up a constant energy shift only, it does not influence the
effective forces between the star and the wall we are mainly 
interested in. The electrostatic-entropic force contributions  
are obtained by differentiating with respect to $D$:
\begin{equation}
\label{eq:deriv}
F_{\rm ee}(D)=-\frac{\partial V_{\rm ee}(D)}{\partial D}
\end{equation}

Now, we derive concrete expressions for the terms of which
${\cal F}_{\rm ee}(D;R,\{N_i\})$ is built up,
whereby we want to include both the counterions' entropic 
contributions and the electrostatic energy. To keep 
our theory as simple as possible, we will omit other thinkable 
contributions, like elastic energies\cite{degennes:book} of the chains
or Flory-type terms arising through 
self-avoidance.\cite{degennes:book,doi:book}.
As we will see 
in Sec.\ \ref{sec:comparison}, our theoretical results in combination with
chain-compression terms to be deduced in what follows, are capable of
producing very good agreement between the theory and
the corresponding simulational data; these simplifications
are therefore {\it a posteriori} justified. Consequently, the free energy
${\cal F}_{\rm ee}(D;R,\{N_i\})$ reads as:
\begin{equation}
\label{eq:helmholtz}
{\cal F}_{\rm ee}(D;R,\{N_i\})=U_{\rm es}+\sum_{i=1}^3 S_i.
\end{equation}

The electrostatic mean-field energy $U_{\rm es}$ is assumed to be given 
by a so-called Hartree-type expression. Let $\rho_{\rm m}({\bf r})$
and $\rho_i({\bf r})$ denote the number densities of the monomers
and the three different counterion species, respectively,
measured with respect to the star's geometrical center. These 
density profiles then determine the overall local charge density  
$\varrho({\bf r})$ of our model system. Therewith, we have:
\begin{equation}
\label{eq:electrostatics}
U_{\rm es}=\frac{1}{2\epsilon}\iint {\rm d}^3r\;{\rm d}^3r'\; 
\frac{\varrho({\bf r})\varrho({\bf r'})}{\left|{\bf r}-{\bf r'}\right|}.
\end{equation}
It is convenient to separate the total charge density $\varrho({\bf r})$
into two contributions: $\varrho_{\rm in}({\bf r})$ in the interior
of the star, i.e., the volume $\Omega_{\rm in}$, and 
$\varrho_{\rm out}({\bf r})$ in the outer region, i.e., the
volume $\Omega_{\rm out}$. Let $\Phi_{\rm in}({\bf r})$ and 
$\Phi_{\rm out}({\bf r})$ be the
contribution to the electrostatic potential at an arbitrary point ${\bf r}$ 
in space caused by the respective charge density. Using this definitions, 
we can rewrite Eq.\ (\ref{eq:electrostatics}) and get
\begin{align}
\label{eq:electrostatics:2}
U_{\rm es}=&\frac{1}{2}\left\{\int_{\Omega_{\rm in}}{\rm d}^3r\;
\left[\Phi_{\rm in}({\bf r})+\Phi_{\rm out}({\bf r})\right]
\varrho_{\rm in}({\bf r})\right.\nonumber\\
&+\left.\int_{\Omega_{\rm out}}{\rm d}^3r\;
\left[\Phi_{\rm in}({\bf r})+\Phi_{\rm out}({\bf r})\right]
\varrho_{\rm out}({\bf r})
\right\}.
\end{align}
On purely dimensional grounds, we expect a result having the 
general form
\begin{align}
\label{eq:electrostatics:3}
\beta U_{\rm es}=&\frac{N_3\lambda_{\rm B}}{R}\cdot h\left(
\frac{R_{\rm W}}{R},\frac{D}{R}\right)\nonumber\\
=&\frac{N_3\lambda_{\rm B}}{R}\left[
h_{\rm in-in}\left(\frac{D}{R}\right)
+2h_{\rm in-out}\left(
\frac{R_{\rm W}}{R},\frac{D}{R}\right)\right.\nonumber\\
&+\left.h_{\rm out-out}\left(
\frac{R_{\rm W}}{R},\frac{D}{R}\right)
\right],
\end{align}
where we introduced dimensionless functions $h_{\alpha-\beta}$ arising
from the integrations of the products 
$\Phi_\alpha({\bf r})\varrho_\beta({\bf r})$ in Eq.\ 
(\ref{eq:electrostatics:2}). Here, one should remember that $N_3$
is the number of uncompensated charges of the star and characterizes
its effective valency. The specific shape of the $h$-functions depends 
on the underlying charge distributions $\varrho_{\rm in}({\bf r})$ 
and $\varrho_{\rm out}({\bf r})$ alone.

The terms $S_i$ represent the ideal entropic free energy contributions 
of the different counterion species. They take the form
\begin{align}
\label{eq:entropy}
\beta S_i=&\int_{\Omega_i}{\rm d}^3r\;\rho_i({\bf r})
\left[\log\left(\rho_i({\bf r})\sigma_{\rm LJ}^3\right)-1\right]
\nonumber\\
&+3N_i\log\left(\frac{\Lambda}{\sigma_{\rm LJ}}\right),
\end{align}
where $\Lambda$ is the thermal de-Broglie wavelength. In writing the sum
in Eq.\ (\ref{eq:helmholtz}), the respective last terms in Eq.\ 
(\ref{eq:entropy}) yield an trivial additive constant only, 
namely $3N_{\rm c}\log(\Lambda/\sigma_{\rm LJ})$, which will be left out
in what follows.

Now, we have to quantitatively address the electrostatic 
and entropic terms pursuant to Eqs.\ (\ref{eq:electrostatics}) to
(\ref{eq:electrostatics:3}) and (\ref{eq:entropy}). For that purpose, 
we first of all need to specify the above introduced number 
densities,\footnote{
Note that, compared to the isolated star case, any influence of the 
wall to the density profiles besides a chopping of the volumes 
available for monomers, monomer ions, and counterions will be omitted.
We will model all densities similar to the approach used in 
Refs.\ \onlinecite{arben:prl:02} and \onlinecite{arben:jcp:02}.}
$\rho_{\rm m}({\bf r})$ and $\rho_i({\bf r})$. Here, we model the arms 
of the PE-star to be fully stretched, or to put
it in other words, the monomer density profile inside the star to 
fall of 
as\cite{klein:macromolecules:99,borisov:jp:97,borisov:epjb:98,arben:prl:02} 
$\rho_{\rm m}({\bf r})=\rho_{\rm m}(r)\sim r^ {-2}$. This 
is a good approximation, as measurements yield a scaling behavior
with an only somewhat smaller exponent $\cong -1.8$, indicating an almost
complete stretching of the chains.\cite{arben:prl:02,arben:jcp:02,
nyquist:macromolecules:32,brilliantov:prl:81,harnau:jcp:112,kantor:prl:83}
Since the monomer ions are placed on the backbone of the chains in a 
periodical manner (cf.\ Sec.\ \ref{sec:md}), their density within the
interior of the star must obviously show an identical $r$-dependence.
Moreover, the profile for the trapped counterions exhibits the
same scaling, due to the system's tendency to achieve local 
electroneutrality.\cite{csajka:macromolecules:33} Therefore, it seems to
be a good choice to use an ansatz $\rho_2(r)=A/r^2$ for the trapped 
counterions and $\varrho_{\rm in}(r)=B/r^2$ for the overall charge density in 
the inner region. Clearly, these distributions have to be normalized 
with respect to the total number of trapped counterions $N_2$ and 
the effective charge of the star $Q^*/e=(N_{\rm c}-N_1-N_2)=N_3$ by 
integrating over the related volumes, $\Omega_2$ and 
$\Omega_{\rm in}$. 
In doing so, we obtain $A=N_2/(2\pi RC)$ and $B=Q^*/(2\pi RC)$, with
\begin{equation}
C=1+\frac{D}{R}
\left[1-\log\left(\frac{D}{R}\right)\right].
\end{equation}

We presume the condensed counterions to be uniformly distributed 
within the tubes surrounding the PE-chains, i.e., we use 
$\rho_1=N_1/\Omega_1$. This approach is supported by simulation results on 
single PE-chains\cite{winkler:prl:80} and was successfully put forward 
in previous studies of PE-star systems.\cite{arben:prl:02,arben:jcp:02}
In a similar fashion, we assume an also uniform distribution of the
free counterions within the outer shell $R<r<R_{\rm W}$, i.e., 
$\rho_3=N_3/\Omega_3$, implying an associated charge density 
$\varrho_{\rm out}=-Q^*/\Omega_3$.

On this basis, we are able to explicitly calculate the variational free 
energy in virtue of Eqs.\ (\ref{eq:helmholtz}) to (\ref{eq:entropy}). 
As far as the entropic contributions are concerned, an analytical 
computation is feasible in a straightforward manner. For 
reasons of clarity, we leave out any intermediate steps and present 
our final results only:
\begin{align}
\label{eq:s1}
\frac{\beta S_1}{N_1}
=&\log\left[\frac{N_1\sigma_{\rm LJ}^3}{\pi(\lambda_{\rm B}^2
-\sigma_{\rm LJ}^2)Rf}\right]-1,\\
\frac{\beta S_2}{N_2}=&\log\left(\frac{N_2}{2\pi C}\right)
+\frac{D}{RC}\log^2\left(\frac{D}{R}\right)
\nonumber\\
\label{eq:s2}
&-3\log\left(\frac{R}{\sigma_{\rm LJ}}\right)+1,\\
\frac{\beta S_3}{N_3}=&\log\left\{\frac{N_3}
{\frac{2\pi}{3}\left[\left(\frac{R_{\rm W}}{R}\right)^3-1\right]
+\frac{\pi D}{R}\left[
\left(\frac{R_{\rm W}}{R}\right)^2+1\right]}\right\}
\nonumber\\
\label{eq:s3}
&-3\log\left(\frac{R}{\sigma_{\rm LJ}}\right)-1.
\end{align}

Now, we want to investigate the electrostatic term $U_{\rm es}$ in more
detail. To begin with, we have to derive expressions for the potential 
functions $\Phi_{\rm in}({\bf r})$ and $\Phi_{\rm out}({\bf r})$. This
computations are rather technical and we will outline the course of
action only. In general, the electrostatic potential $\Phi({\bf r})$ 
due to a charge density $\varrho({\bf r})$ in a dielectric medium 
is given by
\begin{equation}
\Phi({\bf r})=\frac{1}{\epsilon}\int{\rm d}^3r'\;\frac{\varrho({\bf r}')}
{\left|{\bf r}'-{\bf r}\right|}.
\end{equation}
In order to calculate the integral above for a chopped
sphere, yielding the function $\Phi_{\rm in}({\bf r})$, we decompose 
it in infinitesimally thin discs which
are oriented perpendicular to the $z$-axis and cover the whole
sphere. The radii of these discs obviously depend on their position 
with respect to the sphere's center. Afterwards, each disc is object to 
further decomposition into concentric rings. Provided that the charge 
density is spherical symmetric, as is in the case under consideration, 
the charge carried by each ring 
can be easily calculated. The electrostatic potential of a charged 
ring is known from literature,\cite{jackson:book, panofsky:book}
therefore the potential of a disc can be derived by integrating 
over all corresponding rings. An analytical determination of this
integral is possible. The potential of the chopped sphere itself can then
be obtained by another integration over all discs. The latter cannot
be performed analytically, thus one has to resort to a simple, 
one-dimensional numerical integration. For a complete and explicit 
description of the procedure, we refer to Ref.\ \onlinecite{arben:jcp:02},
Appendix A, where an almost identical derivation was put forward.
So far, the potential $\Phi_{\rm out}({\bf r})$ caused by the hollow
chopped sphere of volume $\Omega_{\rm out}$ containing the free counterions
remains to be acquired. Thereto, we employ the superposition principle.
The hollow region of uniform charge density $\varrho_{\rm out}({\bf r})$ 
can be apprehended as a superposition of two chopped spheres with radii
$R$ and $R_{\rm W}$, and charge densities $-\varrho_{\rm out}({\bf r})$
and $\varrho_{\rm out}({\bf r})$, respectively. In doing so, the
problem is reduced to the calculation of the electrostatic potential of 
a chopped sphere with uniform charge density, which can be computed
following the method described above for the inner sphere of volume 
$\Omega_{\rm in}$ (see also Ref.\ \onlinecite{arben:jcp:02}, Appendix B). 
Clearly, with knowledge of both $\Phi_{\rm in}({\bf r})$ and 
$\Phi_{\rm out}({\bf r})$, the dimensionless functions $h_{\alpha-\beta}$ 
and thus the electrostatic energy $U_{\rm es}$ can be obtained according 
to Eqs.\ (\ref{eq:electrostatics:2}) and (\ref{eq:electrostatics:3}) using 
numerical techniques.

In principle, the electrostatic-entropic interaction potential 
$V_{\rm ee}(D)$ is obtained by adding up the entropic and
electrostatic contributions following Eq.\ (\ref{eq:helmholtz}) and
minimizing the Helmholtz free energy ${\cal F}_{\rm ee}(D;R,\{N_i\})$ 
with respect to $R$ and $\{N_i\}$, cf.\ Eq.\ (\ref{eq:potential}). 
However, the star radii are in good approximation unaffected by the
center-to-wall separation $D$, as confirmed by our simulations.
The physical reason lies in the already almost complete stretching
of the chains due to their charging. 
In the simulation the radius $R$ of the star was measured according to 
\begin{equation}
R^2=\frac{1}{f}\left<\sum_{\nu=1}^{f}\left({\bf r}_{\nu N}-{\bf r}_{0}\right)^2\right>,
\end{equation}
where ${\bf r}_{\nu N}$ stands for the position vector of the last monomer
of the $\nu$-th arm of the star and ${\bf r}_0$ the core position. 
Fig.\ \ref{fig:r} illustrates the weak $R$ vs. $D$-dependence for an 
exemplarily chosen parameter combination. Therefore, it is 
convenient not to determine $R$ through the variational calculation, 
but to use average values $\left<R\right>_{D}$ as obtained from
MD simulations instead. The latter are comparable to the 
corresponding radii for isolated PE-stars according to 
Ref.\ \onlinecite{arben:jcp:02}, see Table \ref{table:radii}. 

\begin{figure}[h]
\begin{center}
\includegraphics[width=8cm,draft=false,clip]
{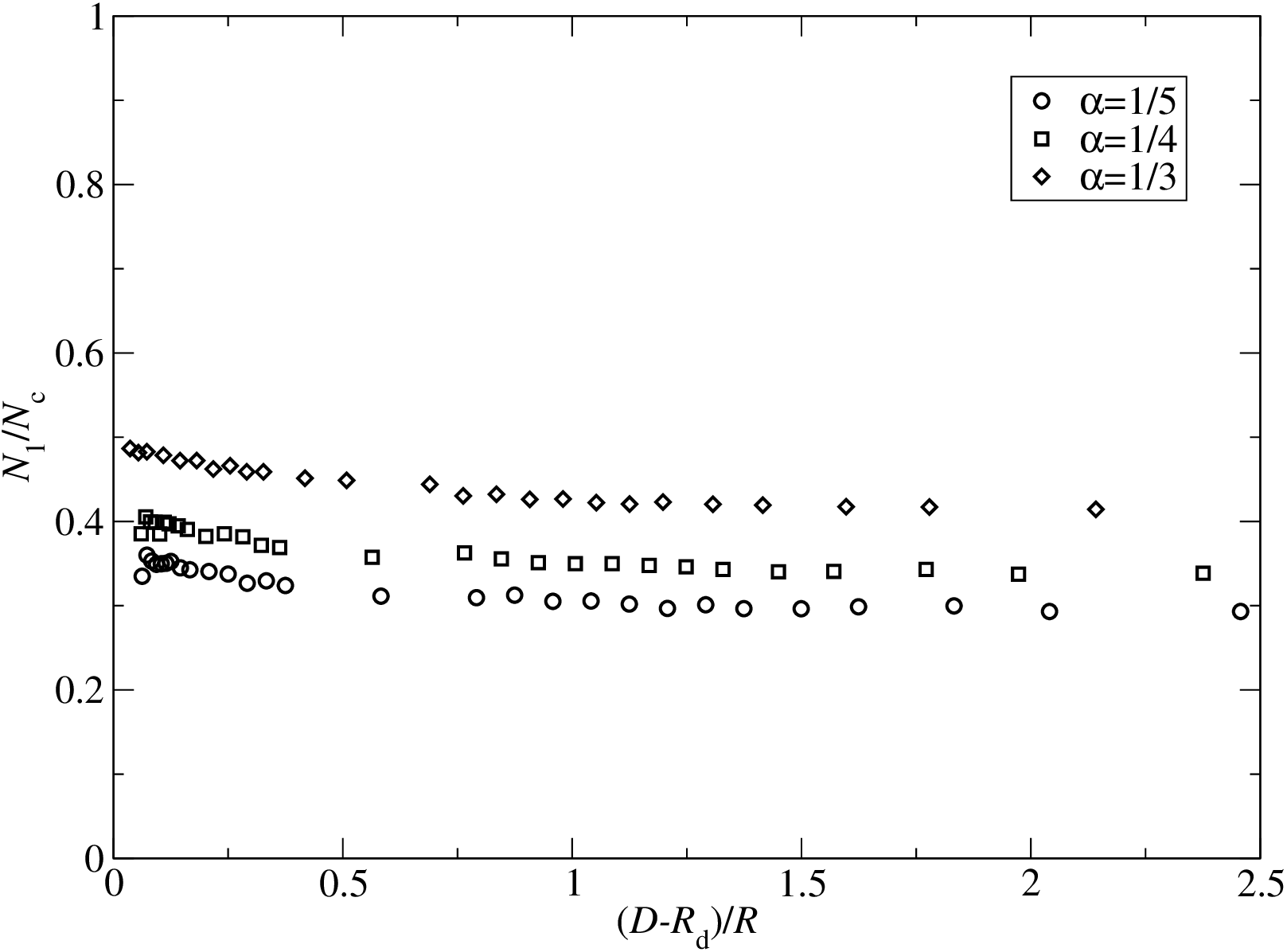}
\includegraphics[width=8cm,draft=false,clip]
{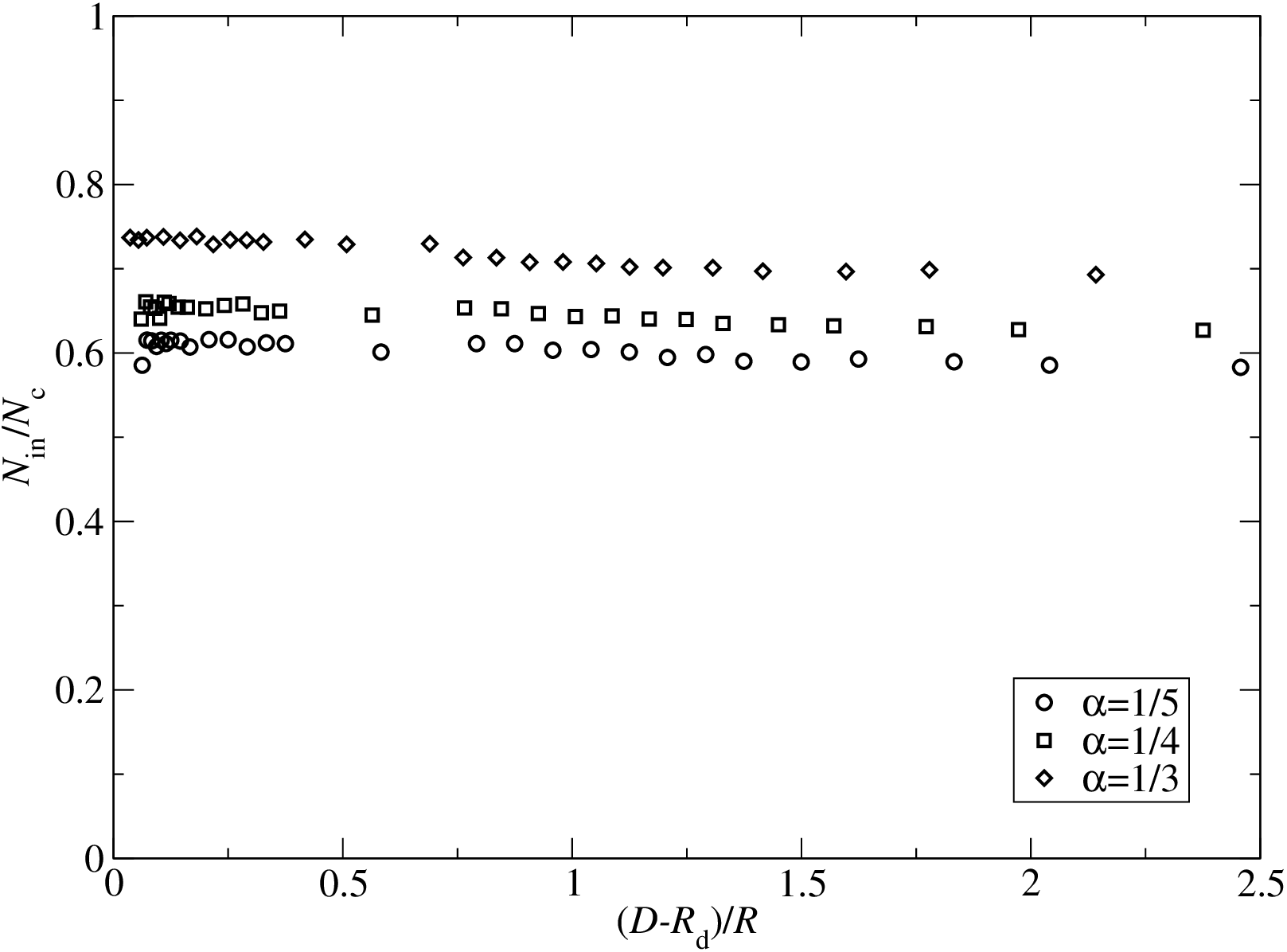}
\end{center}
\vspace{-0.5cm}
\caption{
\label{fig:n-species}
Konieczny, Likos}
\end{figure}

The amount of condensed counterions $N_1$ was measured by counting
the number of such particles separated from the monomer ions by a distance
smaller than $\lambda_{\rm B}$ and performing a statistical average.
The total number of captured counterions, $N_{\rm in}=N_1+N_2$, was 
measured by counting all counterions within a sphere having the 
instantaneous, arm-averaged center-to-end distance and again taking a
time average.
Since the $\{N_i\}$ are related through the equation 
$N_{\rm c}=N_1+N_2+N_3$, only two independent variational parameters 
remain, say $N_1$ and $N_2$. In our simulations, we have found that 
the number of condensed counterions, $N_1$, is also approximately 
constant with respect to $D$, see Fig.\ \ref{fig:n-species}. Hence, 
we will treat $N_1$ as a fit parameter, held constant for all $D$ 
and chosen in such a way to achieve optimal agreement with simulation
results. Therefore, ${\cal F}_{\rm ee}(D;R,\{N_i\})$ is minimized with
respect to the number of trapped counterions, $N_2$, only, reflecting
the possibility of redistribution of counterions between inside/outside
the stars as the distance to the wall is varied. 

Fig.\ \ref{fig:f-contrib} shows a comparison of the entropic and 
the electrostatic contributions $S$ and $U_{\rm es}$ to the 
electrostatic-entropic effective potential, $V_{\rm ee}$. As one can see 
from the exemplary plot, the total entropic term $S$ is the major 
contribution to $V_{\rm ee}$ and therefore determines the functional 
form of the latter, while the electrostatic term is of minor 
importance. At first glance, this may seem to be 
counterintuitive in a PE-system, but electrostatics do in fact 
indirectly affect the interaction potential: Due to the presence 
of charges, the conformations and thus the radii of PE-stars are 
strongly changed compared to the neutral star case, leading to an 
increased range of the interaction. In the inset of 
Fig.\ \ref{fig:f-contrib} it can be seen that total entropy 
of the system at hand is mainly determined by the trapped counterions' 
contribution $S_2$. As one can see, $S_3$ is only weakly influenced 
by the star-to-wall separation $D$, and $S_1$ is even completely 
independent of $D$, as was evident from Eq.\ (\ref{eq:s1}). 
We have to emphasize, however, that even though $S_1$ contributes a constant 
value only and therefore does not influence the effective force
$F_{\rm ee}(D)$ at all, the number of condensed counterions 
$N_1$ nevertheless plays an fundamental role in our problem. Since 
$N_2=N_{\rm in}-N_1$ set the overall scale for of the term $S_2$, 
$N_1$ becomes relevant in renormalizing the effective interaction.
The treatment of $N_1$ and the net charge, 
consequently, as a fit parameter is a common approach for charged 
systems, known as charge renormalization.\cite{hansen:arpc:51,
levin:physicaa:225,tamashiro:physicaa:258}

\subsection{The chain compression contribution}

\label{subsec:cc}

In anticipation of Sec.\ \ref{sec:comparison}, we want to point out
that significant deviations arise between the effective star-wall 
forces as obtained by our computer simulations and their theoretical 
counterparts $F_{\rm ee}$ calculated using the formalism presented
in Sec.\ \ref{subsec:ee} for intermediate center-to-surface 
distances $D/R\approx 0.6\ldots 1.3$. Now, we want to elucidate 
the physical mechanism leading to these deviations and derive 
simple expressions for additional contributions to both the 
total effective forces and potentials, $F_{\rm c}$ and $V_{\rm c}$. 
Therewith, we will conclude the development of our theoretical
approach. 

\begin{figure}[h]
\begin{center}
\includegraphics[width=8cm,draft=false,clip]
{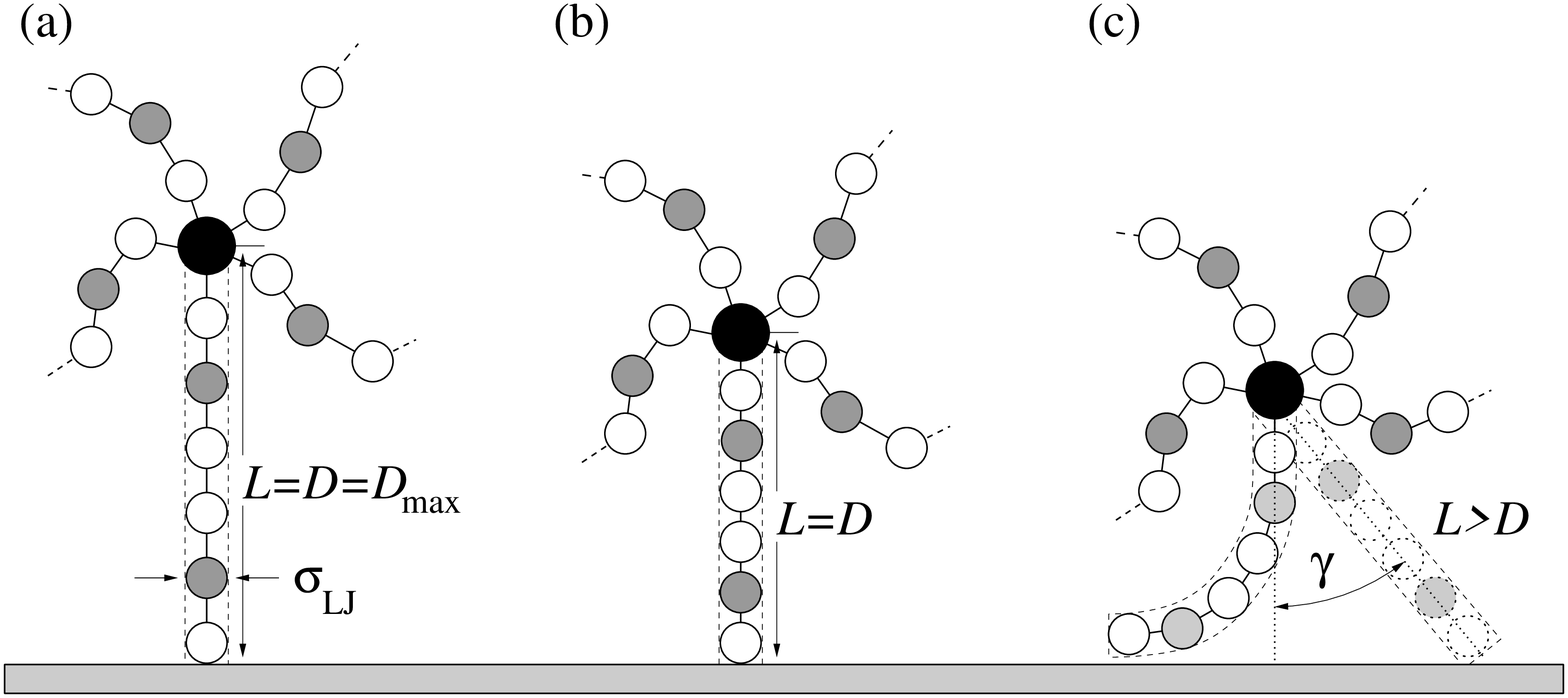}
\end{center}
\vspace{-0.5cm}
\caption{
\label{fig:arm-comp}
Konieczny, Likos
}
\end{figure}

The need to introduce these supplementary contributions is due to so far 
unconsidered conformational changes enforced by the presence of the 
confining wall. During the construction of the mean-field part of our 
theory, we neglected such changes by assuming density profiles of monomers
and counterions which are undisturbed compared to the case of isolated 
stars in full 3d geometries. The influence of the wall became 
manifest in a truncation of the spheres representing star and 
associated counterion cloud, only. But in reality, for $D$
in the order of the typical length of an arm of the star, the star
will undergo strong configurational variations to avoid the wall 
since the monomers are not able to interpenetrate it. Thereby, due 
to the presence of neighboring arms, it can be energetically favorable 
for chains directing towards the wall to compress instead of bending 
away from the surface, although this compression leads to an extra 
cost in electrostatic energy.  The latter obviously originates in 
a decrease of the ion-ion distances along the backbones of the 
chains. Figs.\ \ref{fig:arm-comp}(a) and \ref{fig:arm-comp}(b) depict 
the situation. Once a critical value of the chain length is 
reached, a further shortening of the chains under consideration 
becomes disadvantageous. Electrostatic repulsions and excluded 
volume interaction increase more and more strongly, the chains 
will preferably curve and start to relax in length, see Fig.\ 
\ref{fig:arm-comp}(c). The occurrence of such compression-decompression 
processes is proved by our simulation runs.

In what follows, we will use a simplified picture and model 
the affected arms as rigid, uniformly charged rods of common length 
$L$ and diameter $\sigma_{\rm LJ}$ for both the compressed and the
bent regime. In the former case, we assume the rods to have
orientations perpendicular to the wall surface. In the latter case,
we will account for the bending only via the re-lengthening of the 
rods and their change in orientation with respect to the $z$-axis. 
Fig.\ \ref{fig:arm-comp}(c) visualizes this approximation, whereby 
the shaded straight chain represents the imaginary rod that replaces 
the bent arm within the framework of our modeling. The 
total charge of one such rod is just $qe\alpha N$, 
leading to an $L$-dependent linear charge density, $\eta$,  
given by $\eta=qe\alpha N/L$. 
We consider then the self-energy
of one rod, as a function of its respective length $L$, to be made up
of a purely repulsive contribution which is electrostatic in nature
and arises due to like-charge repulsions, and a second harmonic term
with an effective spring constant $k_{\rm eff}$ describing the binding 
of the chain monomers in a coarse-grained fashion. Thus, we 
have:\cite{nguyen:jcp:114}
\begin{align}
\beta U_{\rm rod}(L)
&=U_{\rm rep}(L)+U_{\rm attr}(L)\nonumber\\
&=\beta\frac{\eta^2}{\epsilon}L\ln
\left(\frac{L}{\sigma_{\rm LJ}}\right)
-\frac{k_{\rm eff}}{2}L^2
\nonumber\\
&=\frac{Z^2\alpha^2N^2\lambda_{\rm B}}{L}\ln
\left(\frac{L}{\sigma_{\rm LJ}}\right)
-\frac{k_{\rm eff}}{2}L^2.
\label{eq:correction:1}
\end{align} 
From Eq.\ (\ref{eq:correction:1}), we then obtain the corresponding
force $F_{\rm rod}(L)=-\partial U_{\rm rod}/\partial L$ that acts on
the ends of the rod parallel to its direction (a negative/positive
sign of $F_{\rm rod}(L)$ corresponds to a force that tends to
compress/stretch the rod.)
The competition of a repulsive and an attractive 
part results in a finite equilibrium length $L_0$ of the rod, i.e.,
$F_{\rm rod}$ vanishes for $L=L_0$. Clearly, $L_0$ is related to the above 
introduced spring constant $k_{\rm eff}$ by the following condition:
\begin{equation}
k_{\rm eff}=-\frac{1}{L_0}
\left.\frac{\partial U_{\rm rep}}{\partial L}\right|_{L_0}.
\end{equation}

Now, let $D_{\rm min}$ and $D_{\rm max}$ determine the range of 
center-to-surface distances for which the above described processes 
take place. Moreover, $D_0$ denotes the critical length for which
the transition from the compressing to the bending regime is observed.
According to our MD results, we will fix $D_{\rm max}/R=1.3$, 
$D_0/R=0.9$, and $D_{\rm min}/R=(2D_0-D_{\rm max})/R=0.5$. In addition,
we require $L_0=D_{\rm max}$ in what follows. We know from our 
simulation runs that it is a reasonable first-order approximation to
assume the length of the affected chains as a function of $D$ to be:
\begin{equation}
L(D)=
\begin{cases}
D_{\rm max} & D\in[0,D_{\rm min}[\\
2D_0-D & D\in[D_{\rm min},D_0[\\
D & D\in [D_0,D_{\rm max}]\\
D_{\rm max} & D\in[D_{\rm max},\infty[.
\end{cases}
\end{equation}
By using this empirical fact, we implicitly include effects due to 
chain bending and entropic repulsions of neighboring chains, even if 
we did not consider corresponding energy contributions explicitly in 
Eq.\ (\ref{eq:correction:1}). Therewith, a promising estimate for the 
chain compression contribution to the total effective force is:
\begin{equation}
F_{\rm c}(D)=
\begin{cases}
f_{\rm eff}\frac{D}{L(D)}
F_{\rm rod}(L(D)) & D\in [D_{\rm min},D_{\rm max}]\\
0 & {\rm otherwise}.
\end{cases}
\label{eq:correction:2}
\end{equation}
Here, $f_{\rm eff}$ is the total number of affected chains. Assuming
that the chains are regularly attached to the colloidal core, we 
expect a linear relation between $f_{\rm eff}$ and $f$, namely
$f_{\rm eff}=f/f_0$. Simulation data indicate $f_0=4$ to be a
good choice for all parameter combinations under investigation.
Note that the pre-factor $D/L(D)=\cos\gamma$
results from geometrical considerations, as can be seen from 
Fig.\ \ref{fig:arm-comp}(c).

Based on Eq.\ (\ref{eq:correction:2}), we obtain the corresponding 
energy term $V_{\rm c}$ by a simple integration:
\begin{equation}
V_{\rm c}(D)=\int_{\infty}^{D}F_{\rm c}(D')\;{\rm d}D'.
\end{equation}

Finally, the total effective forces and interaction potentials
are obtained as the sum of the electrostatic-entropic and
compression terms:
$F_{\rm eff} = F_{\rm ee}+F_{\rm c}$ and 
$V_{\rm eff} = V_{\rm ee}+V_{\rm c}$, 
respectively. Fig.\ \ref{fig:f-correct} 
shows a concluding comparison of the full effective potential $V_{\rm eff}(D)$
and the same without the inclusion of the compression term
$V_{\rm c}(D)$. 
According to the plot, the theoretically predicted 
total potential is purely repulsive and ultra-soft. As we will see in
Sec.\ \ref{sec:comparison}, the total effective forces show a striking 
step-like shape for intermediate distances. This is in contrast to the 
well-known case of star-star interactions.\cite{arben:prl:02,arben:jcp:02}

\section{Comparison and discussion}

\label{sec:comparison}

In this section we test our theoretical model against corresponding 
simulation results to confirm its validity. Using standard MD 
techniques, a straightforward measurement of (effective) interaction 
potentials is not possible. Thereto, one would have to apply more 
sophisticated methods.\cite{watzlawek:book} But since the mean force 
acting on the center of a PE-star can be easily received from computer 
experiments,\cite{jusufi:macromolecules:32,likos:physrep:348} we will 
focus on a comparison of such effective forces and corresponding 
theoretical predictions. To be more precise: when choosing the 
colloidal center of a star at a fixed position ${\bf r}_{\rm core}$ 
as effective coordinate in our simulations, the effective force 
can be measured as the time average over all instantaneous forces 
${\bf f}_{\rm core}$ acting on the core (cf.\ also the simulation 
model in \ Sec.\ \ref{sec:md}) by means of the expression:
\begin{equation}
{\bf F}({\bf r}_{\rm core})
=-\nabla_{{\bf r}_{\rm core}} V_{\rm eff}({\bf r}_{\rm core})
=\left<{\bf f}_{\rm core}\right>.
\end{equation}

In all considerations to come, we will deal with absolute values
of the forces only. It is obvious from the chosen geometry that 
the effective forces are in average directed perpendicular to the
confining wall, i.e., parallel to the $z$-axis. Thus, the mean values
of the $x$- and $y$-components vanish, leading to the relation
$F_{\rm eff}=\left|{\bf F}_{\rm eff}\right|=F_{{\rm eff},z}$. 
For symmetry reasons, the effective forces and potentials depend
on the $z$-component of ${\bf r}_{\rm core}$ alone, 
whereas $r_{{\rm core},z}=D$. Consequently, the forces are connected 
to the potential by the following simple equation: 
\begin{equation}
F_{\rm eff}({\bf r}_{\rm core})
=F_{\rm eff}(D)
=-\frac{\partial V_{\rm eff}}{\partial D}.
\end{equation}
Therewith, predictions for the effective forces can be computed 
starting from theoretical results for the corresponding potential 
(see Sec.\ \ref{sec:theory}, Fig.\ \ref{fig:f-correct}). This 
allows the desired comparison to MD data. 

\begin{figure*}[h]
\begin{center}
\frame{
\rule{0cm}{5cm}
\includegraphics[width=8cm,draft=false,clip]
{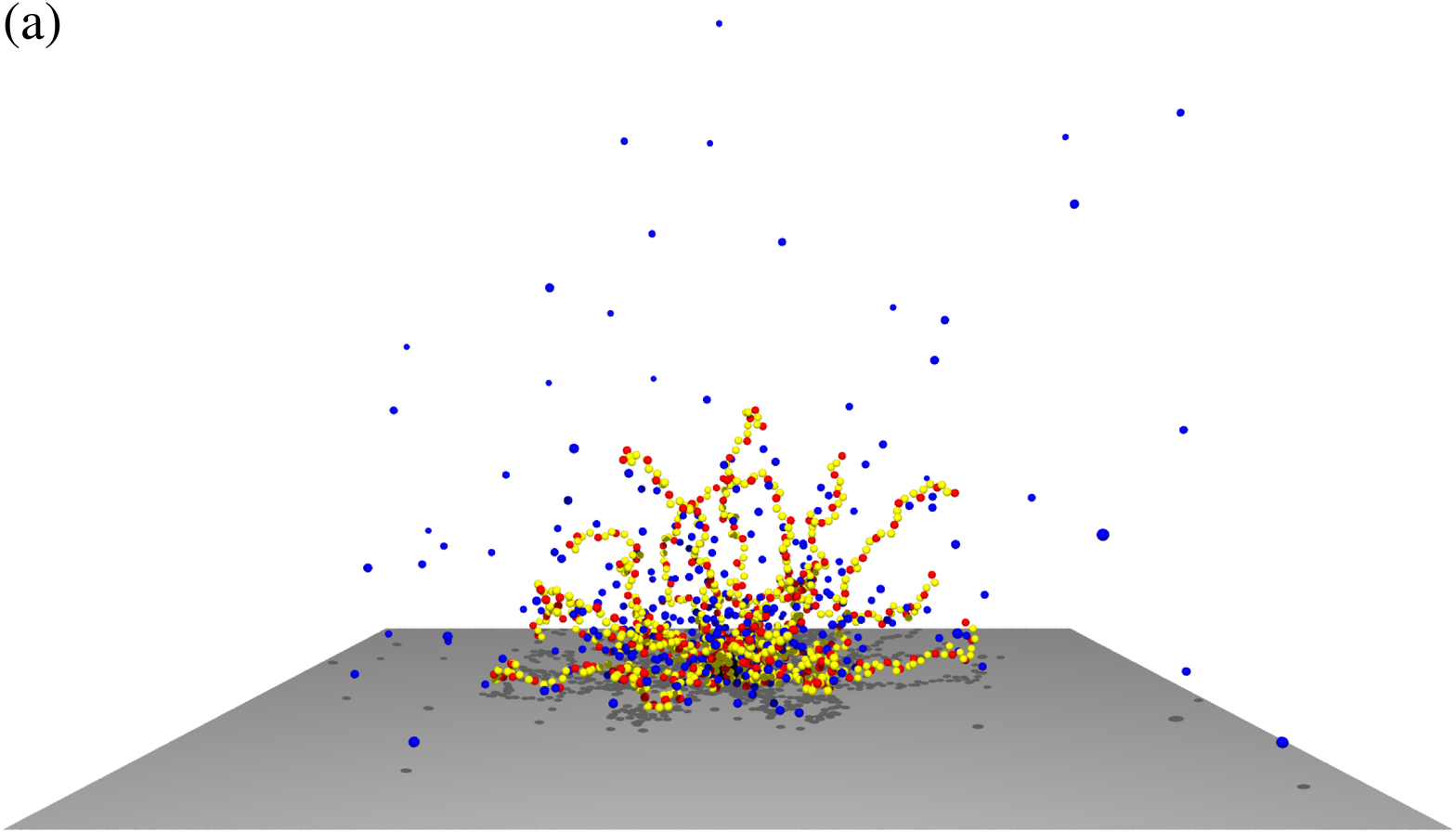}
}
\frame{
\rule{0cm}{5cm}
\includegraphics[width=8cm,draft=false,clip]
{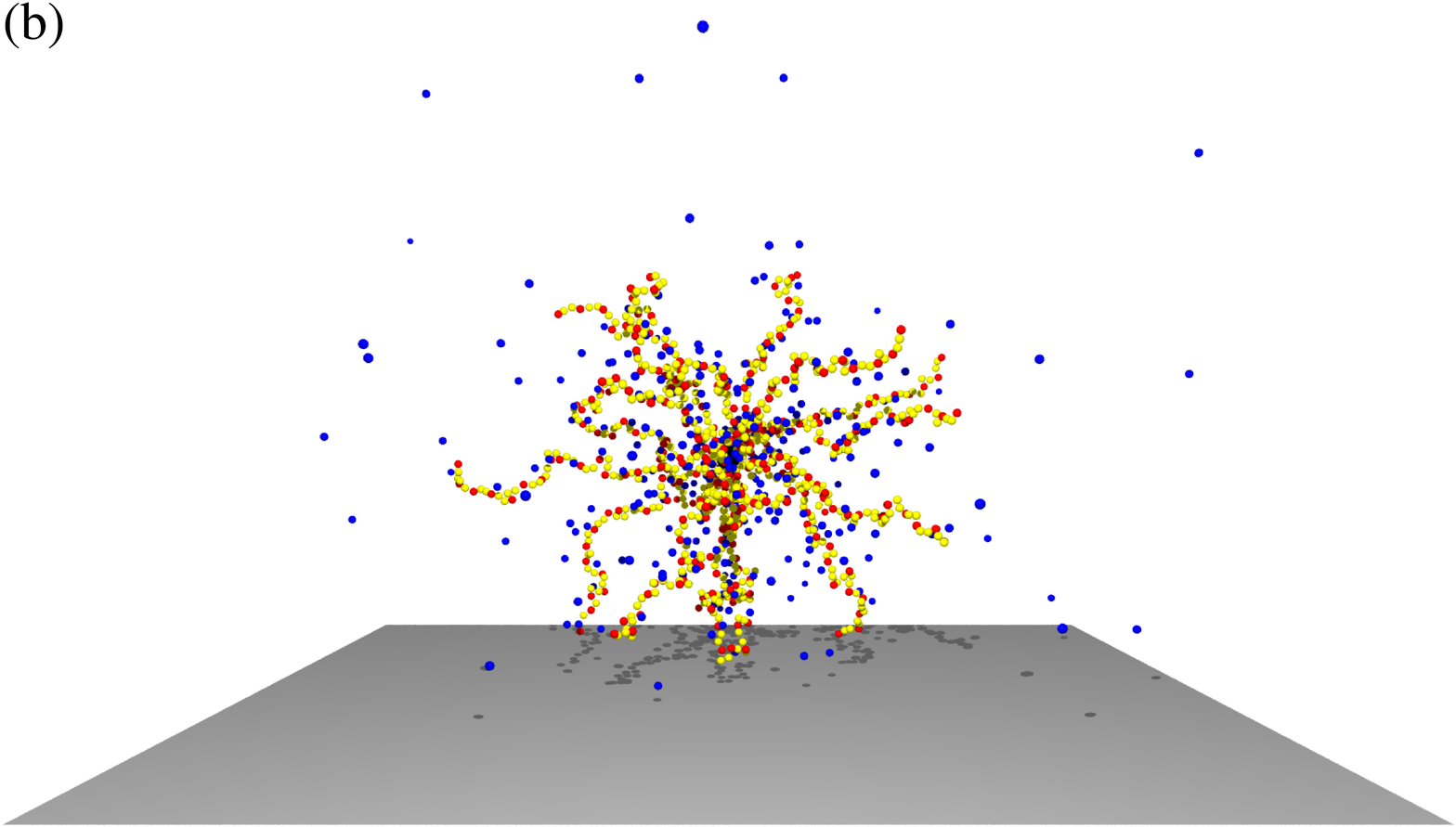}
}
\end{center}
\vspace{-0.5cm}
\caption{
\label{fig:md-snapshot}
Konieczny, Likos
}
\end{figure*}

Fig.\ \ref{fig:md-snapshot} shows typical simulation snapshots for two 
different values of the center-to-surface separation $D$, illustrating 
the conformational changes a star undergoes due to the presence of the 
confining wall. In part (b) of the figure, one may particularly
note the PE-chains directed perpendicular to the wall. These arms 
are object to the compression-decompression mechanism described in 
detail in Sec.\ \ref{subsec:cc} and sketched in Fig.\ \ref{fig:arm-comp}. 
Moreover, the snapshots qualitatively visualize our quantitative finding 
that the majority of counterions is captured within the interior of a 
PE-star, cf.\ Fig.\ \ref{fig:n-species}. We want to point out that, compared 
to the bulk case,\cite{arben:prl:02,arben:jcp:02} the counterion 
behavior in this regard is not significantly changed by introducing 
(hard) walls. In this context, we also refer to Table \ref{table:radii}, 
where the number of condensed counterions $N_1$ as obtained from our MD 
runs is shown for different parameter combinations together with 
corresponding values for isolated PE-stars in unconfined geometries.

Fig.\ \ref{fig:f-results} is the core piece of this work and finally 
presents both our simulational and theoretical results for the effective
star-wall forces. As can be seen from the plots, we find a very good 
agreement for all combinations of arm numbers and charging 
fractions. The theory lines exhibit a remarkable step-like shape for 
intermediate values of the center-to-surface distance $D$ arising due to 
the introduction of the chain compression term $F_{\rm c}$. The latter 
had to be added to the electrostatic-entropic forces $F_{\rm ee}$ 
to account for conformational changes induced by the spatial proximity 
of the star to the planar wall. In doing so, the functional form of 
the theoretically predicted forces accurately renders the results found 
by our computer experiments. When using the respective star's radius as 
basic length scale for the plots, the position of the step is independent
of the parameters considered. In this sense, the effect is universal 
and there is no influence by specific details of the system. 
Note that the distinct kink is an artifact resulting from
the special construction of $F_{\rm c}$, it does not have any physical 
meaning. Again, it should be emphasized that the described behavior is 
in contrast to the well-known star-star 
case.\cite{arben:prl:02,arben:jcp:02}
The difference between both systems mainly lies in the fact that a star 
is strictly forbidden to access the region of the wall, while it 
is allowed to interpenetrate another star-branched macromolecule to a 
certain degree. To put it in other words: a wall is impenetrable, while 
a second star is a diffuse, soft object. Here, we also want to
emphasize that the step is no hysterisis/metastability effect, as
in our MD simulations the center-to-surface separation $D$ is varied 
by pulling the stars away from the wall instead of pushing them towards 
it. This procedure prevents any coincidental trapping of individual 
PE-chains.

The particular values of the fit parameter $N_1$ used to calculate 
the theoretical curves in Fig.\ \ref{fig:f-results} are given in the 
corresponding legend boxes. Obviously, $N_1$ significantly grows
with both the charging fraction $\alpha$ and the arm number $f$, as is 
physically reasonable. Table \ref{table:radii} compares the fit values 
used to their counterparts obtained by our simulations. 
Thereby, one recognizes 
systematic deviations, i.e., the fit values are always slightly off from the 
numbers computed using the MD method. Both quantities are of the order 
of 50-60\% of the total counterion number, for all functionalities $f$ and
charging fractions $\alpha$. The discrepancies emerge due to the 
different meanings of the quantities star radius $R$ and number 
of condensed counterions $N_1$. Within the scope of
the simplified theoretical model, a PE-star is a spherical object 
of well-defined radius, while we observe permanent conformational 
fluctuations of the simulated stars. Thus, the MD radius does not 
identify a sharp boundary, but determines a typical length scale only. 
In this sense, the star can 
be viewed as a fluffy object. Most of the time, there are monomers,
monomer ions and condensed counterions located at positions outside 
the imaginary sphere of radius $R$, and such counterions are not 
counted when measuring $N_1$, thus the theory value typically exceeds
the simulation one. With increasing arm number $f$ and charging fraction
$\alpha$, there is less variation in length of different arms of a star, 
resulting in smaller deviations between the theoretically predicted $N_1$ 
and its MD equivalent. Table \ref{table:radii} confirms this trend.

A last remark pertains to the behavior for small center-to-surface 
distances $D$. Since our simulation model includes a colloidal core 
of finite radius $R_{\rm d}$, the forces obtained from MD runs 
diverge for vanishing wall separations due to the core-wall LJ 
repulsion which mimics the excluded volume interaction. In contrast, 
our theoretical model has no core, for this reason the predicted 
forces remain finite for the whole range of distances. According to
this, our theoretical approach is in principle not capable of 
reproducing a divergence of the forces, even in case of close 
proximity to the confining wall. 

\section{Conclusions and outlook}
\label{sec:conclusion}

We have measured by means of Molecular Dynamics simulations and
analyzed theoretically the effective forces emerging when multiarm
polyelectrolyte stars approach neutral, impenetrable walls. The
forces have the typical range of the star radius, since osmotic
PE-stars reabsorb the vast majority of the counterions and are
thereby almost electroneutral; longer-range forces that could
arise due to the deformation of the diffuse counterion layer
outside the corona radius are very weak, due to the small population
of the free counterion species. The dominant mechanism that
gives rise to the soft repulsion is the entropy of the absorbed
counterions and the reduction of the space available to them
due to the presence of the impenetrable wall. At the same time,
a novel, additional mechanism is at work, which stems from the
compression of a fraction of the star chains against the hard
wall. For star-wall separations that are not too different from
the equilibrium star radius, the chain compression gives rise
to an additional repulsive contribution to the force. At close
star-wall approaches, the compressed chains `slip away' to the
side, orient themselves parallel to or away from the wall and thus
the compression process ceases and the additional contribution
to the force vanishes.

The compression force could play a decisive role in influencing
the cross-interaction between PE-stars and spherical, hard colloids
of larger diameter. Indeed, for this case, the cross-interaction
could be calculated by using the results of the present work as
a starting point and performing a Derjaguin approximation. It will
be interesting to see what effects the cross-interaction has in
the structural and phase behavior of such PE-star--colloidal mixtures.
This issue, along with the study of the 
effects of charged walls and wall-particle dispersion forces,
are the subject of ongoing work. 

\begin{acknowledgments}
The authors wish to thank Prof.\ L.\ Andrew Lyon (Georgia Tech) for 
bringing Refs.\ \onlinecite{kim:jacs:04,serpe:am:04,kim:ac:05,kim:jacs:05}
to their attention.
\end{acknowledgments}



\clearpage

\begin{center}
TABLES
\end{center}

\begin{table}[h]
\caption{
\label{table:radii}
Conformational properties as obtained from MD simulations and 
corresponding fit parameters used in our theoretical approach for
different arm numbers and charging fractions. In addition, results
from Ref.\ \onlinecite{arben:jcp:02} are presented for comparison.
The chain length is fixed to $N=50$. For our data, the Wigner-Seitz 
radius is $R_{\rm W}/\sigma_{\rm LJ}=74.44$. Note that there are
in part insignificant discrepancies between the parameters used here
and in Ref.\ \onlinecite{arben:jcp:02}.}
\begin{ruledtabular}
\begin{tabular}{cccccccc}
$f$ & $\alpha$ & $N_{\rm c}$ & 
$(R/\sigma_{\rm LJ})$\footnotemark[1] & 
$(N_1)$\footnotemark[1] & 
$(N_1)$\footnotemark[2] & 
$(R/\sigma_{\rm LJ})$\footnotemark[3] & 
$(N_1)$\footnotemark[3]\\
\hline
10 & 1/5 & 100 & 24.0 & 32  & 53  & -    & -    \\
10 & 1/4 & 120 & 24.8 & 44  & 68  & 25.3 & 46   \\ 
10 & 1/3 & 170 & 27.5 & 76  & 107 & 27.4 & 72   \\
\hline
18 & 1/5 & 180 & 24.6 & 78  & 80  & -    & -    \\
18 & 1/4 & 216 & 25.4 & 104 & 111 & 26.6 & 107  \\
18 & 1/3 & 306 & 28.0 & 163 & 183 & 28.3 & 159  \\
\hline
30 & 1/5 & 300 & 25.1 & 161 & 113 & -    & -    \\
30 & 1/4 & 360 & 25.9 & 208 & 170 & 27.2 & 213  \\
30 & 1/3 & 510 & 28.4 & 315 & 294 & 28.6 & 309  
\end{tabular}
\end{ruledtabular}
\footnotetext[1]{Values as obtained from our MD simulations, averaged
with respect to $D$. Cf.\ Sec.\ \ref{sec:theory}.}
\footnotetext[2]{Fit parameter used when calculating theoretical 
predictions for star-wall forces.}
\footnotetext[3]{Simulation results for isolated PE-stars,
taken from Ref.\ \onlinecite{arben:jcp:02}, shown for comparison.}
\end{table}


\clearpage

\begin{center}
FIGURE CAPTIONS
\end{center}

FIG.\ 1: Sketch visualizing the physical situation at hand, namely showing 
a PE-star of typical spatial extent $R$, and its counterion background of 
radius $R_W$. The three possible counterion states are illustrated: Condensed, 
trapped, and free. For further explanations, see main text.

FIG.\ 2: Star radii vs.\ the star-wall separation $D$ as obtained
by MD simulations, referring to parameters $f=10$, $\alpha=1/3$. The mean 
value, averaged with respect to $D$, is shown as a horizontal line 
for comparison. 

FIG.\ 3: Fractions of condensed counterions $N_1/N_{\rm c}$ and captured counterions
$N_{\rm in}/N_{\rm c}=(N_1+N_2)/N_{\rm c}$ for PE-stars with $f=10$ arms 
and various values of $\alpha$. As one can see from the plots, both 
quantities are in very good approximation $D$-independent and the majority
of counterions is captured within the interior of the star.

FIG.\ 4: Comparison of the entropic and the electrostatic contributions $S$
and $U_{\rm es}$ to the electrostatic-entropic part of 
the effective potential, $V_{\rm ee}$, exemplarily 
shown for $f=10$, $\alpha=1/3$, and $N_1=107$. Obviously, the entropic 
contribution dominantly determines the $D$-dependence of $V_{\rm ee}$. 
In the inset, $S$ is decomposed into the different counterion species' 
contributions $S_i$, illustrating that the term $S_2$ governs the 
functional form of the total entropy. Note that the different contributions 
in both the main plot and the inset were shifted by constant values to
enhance the readability of the plot. 

FIG.\ 5: Schematic illustrating the physical mechanism leading to the 
necessity to introduce the energy contribution $V_{\rm c}$. Hollow circles 
denote neutral monomers, gray balls are monomer ions. The bigger black ball 
is the colloidal core of the star. Counterions are omitted in the 
depiction for reasons of clarity. In part (c) of the figure, the dashed 
rod replacing the chain bent to the left is shown mirror-reflected to 
the right to avoid crowding. For a more detailed discussion of the 
compression effect, see main text. 

FIG.\ 6: Comparison of the total and the electrostatic-entropic 
contributions to the effective potential, $V_{\rm eff}(D)$ and 
$V_{\rm ee}(D)$, for $f=10$, $\alpha=1/3$, and $N_1=107$. The inset 
shows the functional form of the compression term $V_{\rm c}(D)$. The 
main text provides an explanation for the need of the latter contribution 
and contains an explicit derivation.

FIG.\ 7: Simulation snapshots of a PE-star with functionality $f=18$ and 
charging fraction $\alpha=1/3$. The center-to-wall distances are 
$D/R=0.107$ (a) and $D/R=1.07$ (b). 
Bright gray balls are neutral monomers, while the dark spheres along the 
chains indicate monomer ions. The counterions are the small, dark gray 
spheres surrounding the star.

FIG.\ 8: Effective star-wall forces with the colloidal core of the respective
stars taken as effective coordinate, obtained from both computer
simulations and our theoretical approach. The fit parameters $N_1$ 
used are specified in the legend boxes. Here, we show results for
stars with functionalities $f=10$ (first row), $f=18$ (second row), and
$f=30$ (third row). In all cases, we are considering charging fractions
$\alpha=1/5$ (first column), $\alpha=1/4$ (second column), and 
$\alpha=1/3$ (third column). In the rightmost picture of the first row,
we additionally included a theoretic line calculated without taking the 
compression term $V_{\rm c}$ into account to illustrate the need of
such a contribution. Since the theoretical model has, in contrast
to the simulation model, no (hard) colloidal core, all simulation data 
have to be displaced by the core radius $R_{\rm d}$.


\clearpage

\clearpage

\begin{figure}[h]
\begin{center}
\includegraphics[width=8cm,draft=false,clip]
{./506620JCP.FIG2.eps}
\end{center}
\vspace{-0.5cm}
\caption{
\label{fig:r}
Konieczny, Likos}
\end{figure}

\clearpage

\clearpage

\begin{figure}[h]
\begin{center}
\includegraphics[width=8cm,draft=false,clip]
{./506620JCP.FIG4.eps}
\end{center}
\vspace{-0.5cm}
\caption{
\label{fig:f-contrib}
Konieczny, Likos
}
\end{figure}

\clearpage

\clearpage

\begin{figure}[h]
\begin{center}
\includegraphics[width=8cm,draft=false,clip]
{./506620JCP.FIG6.eps}
\end{center}
\vspace{-0.5cm}
\caption{
\label{fig:f-correct}
Konieczny, Likos
}
\end{figure}

\clearpage

\clearpage

\begin{figure*}[t]
\begin{center}
\includegraphics[width=5.4cm,draft=false,clip]
{./506620JCP.FIG8a.eps}
\includegraphics[width=5.4cm,draft=false,clip]
{./506620JCP.FIG8b.eps}
\includegraphics[width=5.4cm,draft=false,clip]
{./506620JCP.FIG8c.eps}
\includegraphics[width=5.4cm,draft=false,clip]
{./506620JCP.FIG8d.eps}
\includegraphics[width=5.4cm,draft=false,clip]
{./506620JCP.FIG8e.eps}
\includegraphics[width=5.4cm,draft=false,clip]
{./506620JCP.FIG8f.eps}
\includegraphics[width=5.4cm,draft=false,clip]
{./506620JCP.FIG8g.eps}
\includegraphics[width=5.4cm,draft=false,clip]
{./506620JCP.FIG8h.eps}
\includegraphics[width=5.4cm,draft=false,clip]
{./506620JCP.FIG8i.eps}
\end{center}
\vspace{-0.5cm}
\caption{
\label{fig:f-results}
Konieczny, Likos
}
\end{figure*}

\end{document}